# Measurement of amplitude of the moiré patterns in digital autostereoscopic 3D display

## VLADIMIR SAVELJEV[1,2*] AND SUNG-KYU KIM[2]

[1]*NEMO Lab. Department of Physics, Myongji University, Yongin-si, Gyeonggi-do 17058, Republic of Korea*
[2]*Center for Image Media Research, Korea Institute of Science and Technology, Seoul 136-791, Republic of Korea*
*\*saveljev.vv@gmail.com*

**Abstract:** The article presents the experimental measurements of the amplitude of the moiré patterns in a digital autostereoscopic barrier-type 3D display across a wide angular range with a small increment. The period and orientation of the moiré patterns were also measured as functions of the angle. Simultaneous branches are observed and analyzed. The theoretical interpretation is also given. The results can help preventing or minimizing the moiré effect in displays.

**OCIS codes:** (120.4120) Moiré techniques; (120.2650) Fringe analysis; (110.3000) Image quality assessment.

**1. Introduction**

The moiré effect is caused by the interaction of the rays passing through two or more periodic layers [1, 2]. In digital three-dimensional (3D) autostereoscopic displays based on LCD panels, the layers (the LCD pixel grid and the barrier plate) are highly periodic. The periodicity of both layers is one reason for the periodic moiré patterns to appear in such displays as a result of the interference between the pixels and the barrier.

We consider the moiré effect in three-dimensional (3D) displays with the horizontal parallax (HPO), i.e. in displays with a one-dimensional parallax barrier plate consisting of repeated parallel lines [3, 4]. In order to improve the image quality, the moiré patterns must be eliminated from displays [5 - 7]. On this way, the information about the amplitude is valuable.

In the current research, we consider the amplitude of the moiré patterns in two parallel layers without a gap between them, so as the distance has no effect on the visual appearance of the patterns. A typical period of the optical moiré patterns is measured in millimeters or centimeters. The period and the amplitude characterize any periodic function. The orientation is an additional attribute of a two-dimensional (2D) periodic function.

The periodic layers can be described by their wavevectors. We may assume that there is freedom for coplanar layers to be turned around the normal axis. When the angle between the wavevectors changes, the moiré patterns also change, and correspondingly, all characteristics of the moiré patterns become functions of the turn-around angle. The functions characterizing the moiré patterns are not necessarily monotonous; rather they have local extrema across the angular region. The maxima of the amplitude correspond to the strong patterns, the minima – to the weak patterns or to their absence. We will refer to the moiré period depending on the rotation angle as the period function; similarly, we will briefly call two other functions "the orientation" and "the amplitude".

In the current content, we call the angles of the local maxima "the moiré angles". It was shown previously that the moiré angles in the digital displays with square pixels are rational angles [8] (the tangent of a rational angle is a rational number [9]). Therefore, in respect to the moiré effect in digital displays, the rational angles have special significance, and sometimes we will mention the rational angle by the fraction which means the tangent.

The necessary angular range can be defined by the symmetry of the layout of gratings. In the current case of square pixels and a one-dimensional (1D) linear barrier plate, it is (0, 45°). In order to include the "wings" of the functions and probably non-square pixels, the practical angular range is widened to (-10°, +100°).

The effect of the modulation transfer function (MTF) takes place in displays, when the moiré patterns are observed from a long distance which does not allow our eyes to resolve the individual screen pixels. This is am improved concept of the visibility circle which models the human visual system by a binary function. Such screens are nowadays typical for regular displays, where the pixels are not recognizable as small squares.

To eliminate or reduce the influence of other parameters which may affect the amplitude, in our experiments, the observation direction is always orthogonal to the screen; each photograph is individually calibrated, etc.

At many angles, none or only one pattern is detected. However, at some angles, several patterns can be recognized in parallel. We call them "branches". The independent branches have different orientations. On the other hand, the harmonics depend on each other, i.e., they have multiple periods and are oriented at the same angle. Switched branches were observed in [10]. Simultaneous branches are considered in the current paper. Typically, one of the branches visually prevails and suppresses others, but we used to keep two or three strongest branches with the highest amplitude and the longest period.

The measurements of the period and the orientation of the moiré patterns in 3D displays were made previously [11], although across a limited range and without details near the moiré angles. In the research [12], a relatively large increment of 5° was used. However, to the authors' knowledge, one of few attempts to measure the amplitude of the moiré patterns experimentally is Ref. [13]. It appears that the amplitude of the moiré patterns is more difficult to measure than the period or the orientation. Therefore, a detailed picture of the behavior of the moiré patterns is yet uncertain. As a result, it is still problematic to draw a qualitative comparison of the visual quality of displays in relation to the moiré effect. Therefore, in this paper, we highlight the important role of the estimated and measured amplitude. Such direct measurements were not performed before, as far as we know. Our current paper is an attempt to fill a gap with the lack of the moiré amplitude in digital autostereoscopic displays.

The paper is organized as follows. In Sec. 2 we review a theory of the period and orientation based on the wavevectors of the gratings and the spectral trajectories. Also, we propose the amplitude theory based on the MTF of the human eye, which, in our particular case, can be approximated by polynomials. In Sec. 3, we describe the measurement technique and present the experimental results of measurements of three main parameters of patterns (period, amplitude, and orientation) across a wide angular range with the angular increment about 1°, and even less near the rational angles. Section 4 finalizes the paper.

## 2. Theory

In our layout, the gratings are coplanar; one grid (2D LCD pixels) is static, while another (1D linear barrier) can be rotated, see Fig. 1. In such configuration, the moiré patterns are periodic parallel bands.

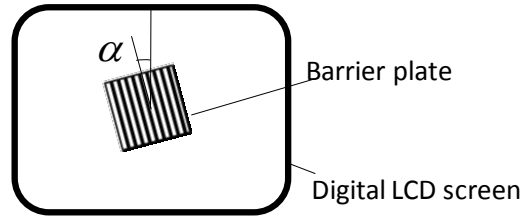

Fig. 1. Layout of coplanar layers.

We imply the square pixels in the theory, although the practical design of LCD displays may deviate from that.

### 2.1 Wavevector, period and orientation of moiré patterns

Generally, the wavevector of a moiré pattern (briefly, the moiré wavevector) is the difference of the wavevectors of the gratings [1]. When the plate rotates, its wavevector (together with all its spectral components) turns around each spectral component of the grid. Among all resulting spectral components, we choose those with trajectories close to the origin of the spectral domain or at least fall within the visibility circle. The extremely important concept of the visibility circle is presented in [1]. In the current paper, the radius of the visibility circle was taken approx. 0.8 of the fundamental wavenumbers; such value ensures the visibility of the moiré patterns, while both pixel grid and the barrier remain invisible.

In the current subsection, instead of the period of the moiré patterns, we prefer using the moiré magnification factor. This moiré factor is defined as the ratio of the moiré period to the period of the grating. For the concept of the moiré magnifier, refer to [16].

Consider the 2D spectrum of the square pixel grid, where $k_0$ is the first spectral component of the spectrum, i.e., the fundamental wavenumber. The spectral components are located at the nodes with integer coordinates $mk_0$ and $nk_0$ ($m$, $n$ are integer numbers). In the complex plane, the center of the trajectory is located at $z_c = (m + jn)k_0$, where $j$ is the imaginary unit; as in Refs. [14], [15], and [17], where the complex numbers are also used. Figure 2 shows an example of one spectral component with arbitrary chosen coordinates 5 and 2, which is combined with the rotated plate. As a result, a spectral trajectory appears.

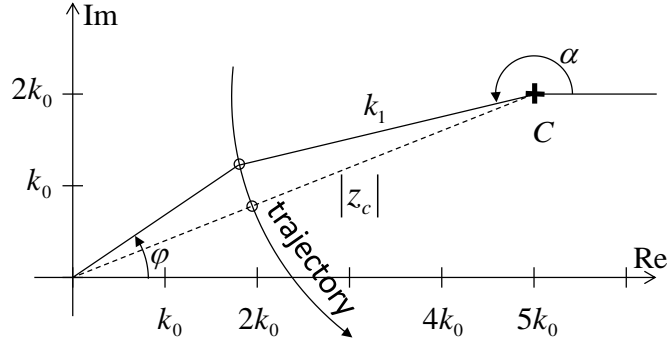

Fig. 2. An example trajectory in the complex plane. The location of the center and the radius are chosen arbitrary.

The moiré wavevector is the ray from the origin to the spectral component of the rotated plate whose center of rotation is at the point $C$. The moiré wavenumber is the distance between the current point on the trajectory and the origin of the spectral domain. The equation of the trajectory with the radius $k_1$ is

$$z = z_c - k_1 \cdot e^{j\alpha} \qquad (1)$$

The complex number $z$ describes the wavevector of the moiré pattern. It can be rewritten in the polar form $z = a \cdot \exp(j\varphi)$, where the modulus $a$ and argument $\varphi$ are the wavenumber and the orientation of the moiré pattern, resp. Namely, the wavenumber of the moiré pattern is as follows (see also Eq. (12) in [17]),

$$k_m = k_1 \sqrt{1 + \rho^2 - 2\rho \cos(\alpha - \alpha_{MAX})} \qquad (2)$$

where $\rho = |z_c|/k_1 = (m^2 + n^2)^{1/2} k_0/k_1$ is the ratio of the periods of the layers, $\alpha_{MAX}$ is the moiré maximum angle which corresponds to the smallest wavenumber and is defined as

$$\alpha_{MAX} = \arctan \frac{n}{m} \qquad (3)$$

The orientation of the pattern is given by

$$\tan \varphi = \frac{nk_0 - k_1 \sin \alpha}{mk_0 - k_1 \cos \alpha} = \frac{\frac{n}{\sqrt{m^2 + n^2}} \rho - \sin \alpha}{\frac{m}{\sqrt{m^2 + n^2}} \rho - \cos \alpha} \qquad (4)$$

Particularly, for identical gratings with the horizontal wavevector (the lines are parallel to the vertical axis, so as $n = 0$, $m = 1$, and $k_1 = 1$), the orientation is given by $\tan \varphi = (\tan \alpha/2)^{-1}$, as known from the book [1].

The period of the moiré patterns is the inverse of the wavevector Eq. (2); it is known in the literature, see, e.g., Eq. (7) in [18]. The corresponding magnification factor is

$$\mu = \frac{1}{\sqrt{1+\rho^2 - 2\rho\cos(\alpha - \alpha_{MAX})}} \tag{5}$$

The theoretical functions defined by Eqs. (5) and (4) are shown in Fig. 3 for different values of $\rho$.

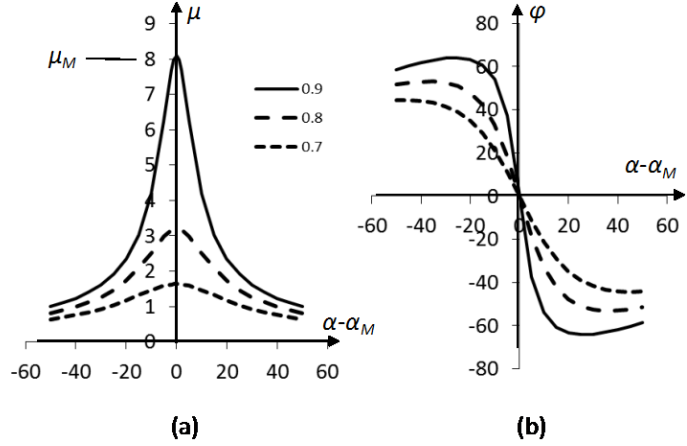

Fig. 3. Theoretical (a) magnification factor and (b) orientation for $\rho = 0.7 \ldots 0.9$.

The graphs Fig. 3 depict the behavior of the patterns near the moiré maximum angle. At that angle, the orientation of the pattern coincides with the orientation of the plate (up to a possible difference in $\pi$ radians),

$$\varphi_{MAX} = \alpha_{MAX} \tag{6}$$

while the maximum moiré factor is as follows (see also Eq. (2-11) in [1]),

$$\mu_{MAX} = \frac{1}{\rho - 1} \tag{7}$$

Equation (7) shows that the period of the moiré patterns is inversely proportional to the difference between the modulus of the spectral component of the grid and the radius of the trajectory. Therefore, the closer is $\rho$ to 1 (or, equivalently $k_1$ to $|z_c| = (m^2 + n^2)^{1/2} k_0$), the longer is the period, see Fig. 3(a); and in the case of $\rho = 1$, it is infinite.

When the rotation angle of the plate approaches the moiré maximum angle, the period of the moiré patterns increases; at the same time, the direction of the moiré wavevector becomes closer to the direction of the wavevector of the plate [11]. At the moiré maximum angle, the period reaches a local maximum, and the wavevector of the patterns is directed along the wavevector of the plate. In this case, the patterns are visually parallel to the lines of the barrier plate; this condition can be used as an indicator of the maximum moiré.

### 2.2 Amplitude of moiré patterns

In a straightforward approach, the amplitude of the spectral components of the moiré patterns within the visibility circle does not depend on the angle of rotation of the plate, since the relations between the amplitudes of the spectral components are not changed due to the rotation.

There are many factors affecting the amplitude of the moiré patterns, and MTF is only one of them. The famous and fruitful concept of the visibility circle is a binary approximation of the MTF. Also, the visibility circle perfectly works in the flat (coplanar) case, but it needs improvement in the spatial case. We suggest using the MTF, for which, the visibility circle is an approximation, see Fig. 4.

The MTF of the human eye (or, equivalently, of the camera) depends on the spatial frequency. Therefore, the visible amplitude of the same spectral component varies depending on the distance from the origin (along the trajectory, different distances correspond to the different angles).

As compared to the previously known angular independence of the amplitude of the moiré patterns derived from the spectral approach, we consider the MTF as a next order approximation to the real display technology. These two cases (visibility circle and MTF) can be probably qualified as the first and the second order approximations.

According to [19], the general formula for the radial MTF of humans is as follows

$$M(u,d,\lambda) = \sqrt{\frac{2}{\pi}} \frac{\sqrt{\arccos \frac{u}{u_0(d,\lambda)} - \frac{u}{u_0(d,\lambda)}\sqrt{1-\left(\frac{u}{u_0(d,\lambda)}\right)^2}}}{\left[1+\left(\frac{u}{u_1(d)}\right)^2\right]^{0.62}} \qquad (8)$$

where $d$ is the diameter of the pupil (2 mm ≤ $d$ ≤ 6 mm), $u$ is the spatial frequency of the patterns, $\lambda$ is the wavelength of the light and

$$u_0(d,\lambda) = \frac{d}{\lambda} \qquad (9)$$

$$u_1(d) = 21.95 - 5.512d + 0.3922d^2 \qquad (10)$$

The spatial frequency of the moiré patterns observed from the distance $L$ is

$$u_M = \frac{L}{\lambda_m \mu} \qquad (11)$$

where $\mu$ is the magnification factor by Eq. (5), $\lambda_m$ the period of moiré fringes on original surface (i.e. the inverse spatial frequency at zero distance from the surface).

For our purpose, it is sufficient to consider the typical parameters $d$ = 4 mm, $\lambda$ = 555 nm. Then $u_0(4, 555)$ = 125.8 and $u_1(4)$ = 6.18. The MTF for these parameters is shown in Fig. 4.

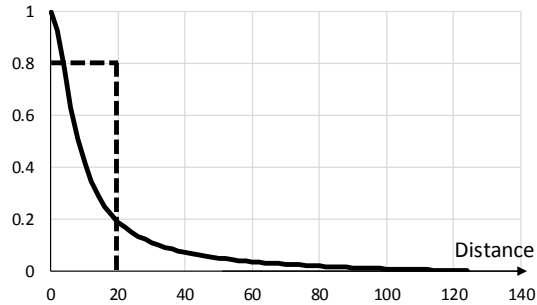

Fig. 4. MTF for $d$ = 4 mm, $\lambda$ = 555 nm. The dashed line shows the profile of the visibility circle.

For the above typical parameters, the MTF Eq. (8) can be approximated, e.g., by the 3rd order polynomial with the interpolation error less than 2% as follows,

$$M(u,4) = -2 \cdot 10^{-6} u^3 + 4 \cdot 10^{-4} u^2 - 3 \cdot 10^{-2} u + 1 \qquad (12)$$

The amplitude calculated basing on Eq. (12) is shown in Fig. 5,

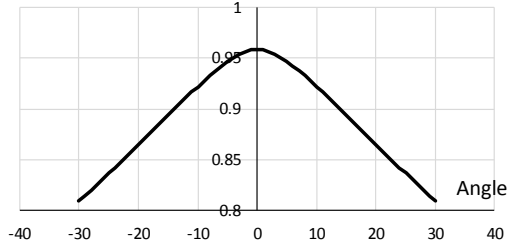

Fig. 5. Theoretical amplitude of the moiré patterns based on MTF.

The shape of the function Fig. 5 near its maximum is close to the triangular function.

## 3. Experiments

In experiments, the barrier plate was placed onto the surface of an LCD screen directly without a gap, and a changeable angle between the axis of the plate and the axis of the screen. For the layout of the experimental setup, refer to Fig. 1.

In experiments, a uniform white field (the pixels of identical brightness across the whole screen) was constantly displayed in an LCD screen, while the plate was turned around. Near specific angles, the moiré patterns appear.

We made experiments with barrier plates of different periods. An experiment with each plate includes many incremental steps. At each step, the plate is turned around by a small angular increment of about 1° (in the neighborhood of the moiré angles, it is smaller), and photographed. As a result, hundreds of experimental photographs of the moiré patterns (technically, plane waves) were taken and processed.

The experiments were made with the barrier samples containing 50, 75, and 150 lines per inch (lpi); these values are often referred to as "pitches". The corresponding periods are 0.508 mm, 0.339 mm, and 0.170 mm resp. The size of each sample is 8.5 cm x 8.5 cm. The period of the screen pixels of the LCD screen is 0.266 mm (the corresponding ratios of periods of the layers $\rho$ are 1.910, 1.273, and 0.634).

In order to provide equal photographic conditions for different plates, several small samples were assembled together on the common base (up to six samples). A photograph of the plate with multiple samples is shown in Fig. 6.

This prevents us from distortions of amplitudes due to possible different illumination conditions of each sample, if photographed individually. For the calibration, we used the samples of longer period, where the intensities in open and close areas could be measured separately. Such assembly allows us to compare the amplitudes in different plates directly.

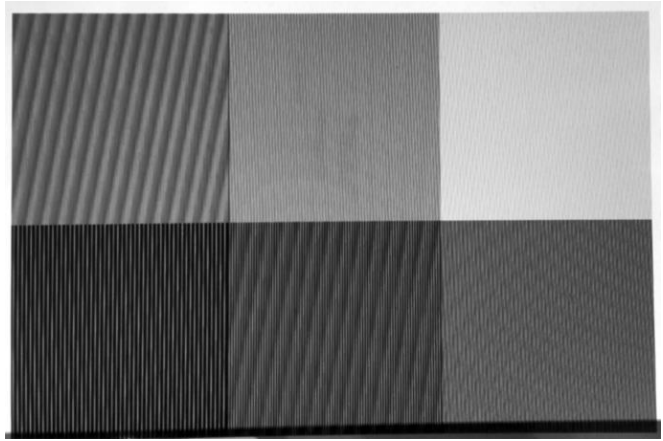

Fig. 6. Experimental photograph of the moiré patterns in the plate with 6 samples (rotation angle = -0.1°).

### 3.1 Image processing

To retrieve the numerical values from numerous photographs, we designed a semi-automatic software system which measures the period, the orientation, and the amplitude of the plain waves in photographs.

The image processing is based on the windowed 2D Fourier transform. For the angular measurements we applied the Hann window; for the amplitude measurements – the flat-top window. The peak's location in the spectral domain determines the orientation and the wavenumber of the moiré patterns, hence, the period. There are numerous spectral peaks in the 2D Fourier transform of each photograph. Therefore firstly, we eliminate the peaks with the lower amplitudes. Then, the peaks with the lower spatial frequencies were selected; this is because the moiré effect corresponds to the longest period.

The branches were classified manually, based on the similarity of the measured parameters (the period and the angle, sometimes the amplitude) along the branch. This similarity means a relatively smooth change of the measured parameter. The maximum amplitude and the maximum period were calculated for each branch individually. Some weak branches of low amplitude and short period were eliminated, especially when these co-exist with the high amplitude or long period branches which used to suppress weak branches visually. In our method, the amplitude of the first spectral component in the spectrum of the moiré patterns is measured.

The illustrations of the data processing are shown in Figs. 7 and 8.

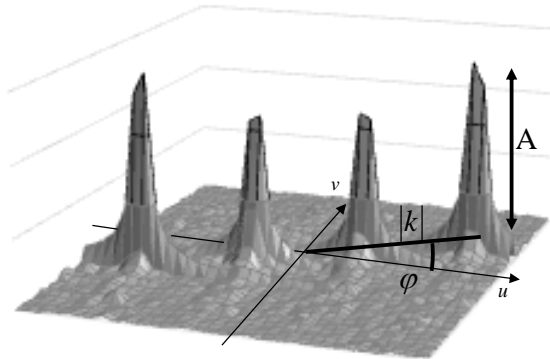

Fig. 7. Measured parameters ($k$, $\varphi$, $A$) of the moiré wave in the spectrum.

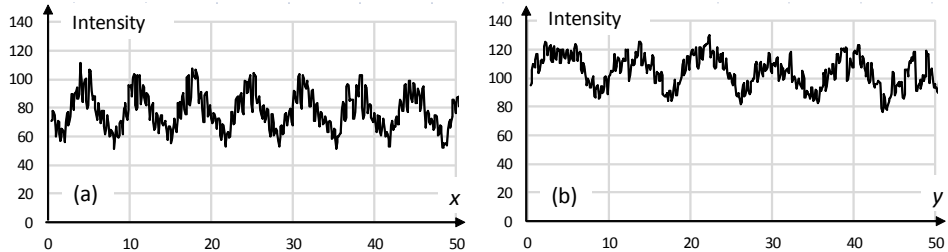

Fig. 8. The profile of the moiré patterns on two orthogonal directions.

Figure 8 shows that the profile of the moiré pattern is essentially non-sinusoidal. This is also an illustration of the practical asymmetry in the horizontal and vertical directions.

*3.2 Measurement of moiré parameters*

Three parameters of the moiré patterns were measured at each rotation angle of the plate: the period, the orientation, and the amplitude. In an example shown in Fig. 9, the patterns were detected experimentally near the angles of 0, 27°, 45°, and 90° which are close to the rational angles 0, 1/2, 1, and ∞. Note that except for the neighborhood of these rational angles, the moiré patterns were not detected in this experiment. For other plates, the angles, where the moiré patterns appear could be different, but nevertheless they are close to other rational angles. Note that Figs. 8 – 11 are obtained experimentally, and therefore they contain some noise.

As described in Sec. 2.1, at the moiré maximum angle, the orientation of the moiré wave coincides with the angle of the plate. Such function has values between -90° and 180°. Instead of it, we consider the modified ("wrapped") function $\varphi - \alpha$, which has symmetrical range of values from -90° to +90°, and the moiré maximum angles lie on the intersections of the function with the abscissa. We prefer this function because of its convenient graphical layout.

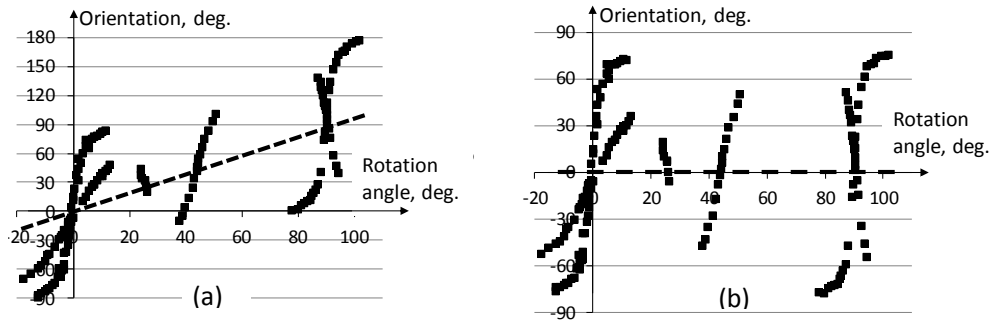

Fig. 9. (a) Measured (unwrap) and (b) modified (wrap) orientation functions based on the experimental data.

Generally speaking, several harmonics could be recognized in a non-sinusoidal wave which is shown, e.g., in Fig. 8. The harmonics are characterized by different periods and smaller amplitudes, but the same orientation. Therefore, the "presumable branches" with the coinciding angle functions and the double or triple periods are not the branches, rather the harmonics as shown in Fig. 10 by square and triangular markers. The higher harmonics are excluded from the measurements, so as the longest period remains.

The examples of three measured parameters are shown in Fig. 11. The amplitude is given in the arbitrary units (after the calibration between the plates by pure black and pure white areas).

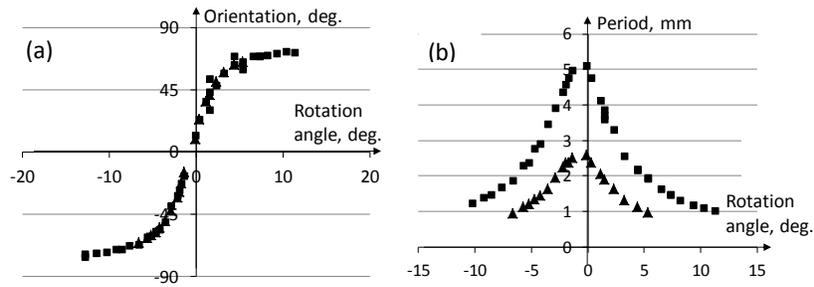

Fig. 10. Two harmonics of a non-sinusoidal moiré wave observed in experiment: (a) orientation (the same for both harmonics); (b) period (differs twice);

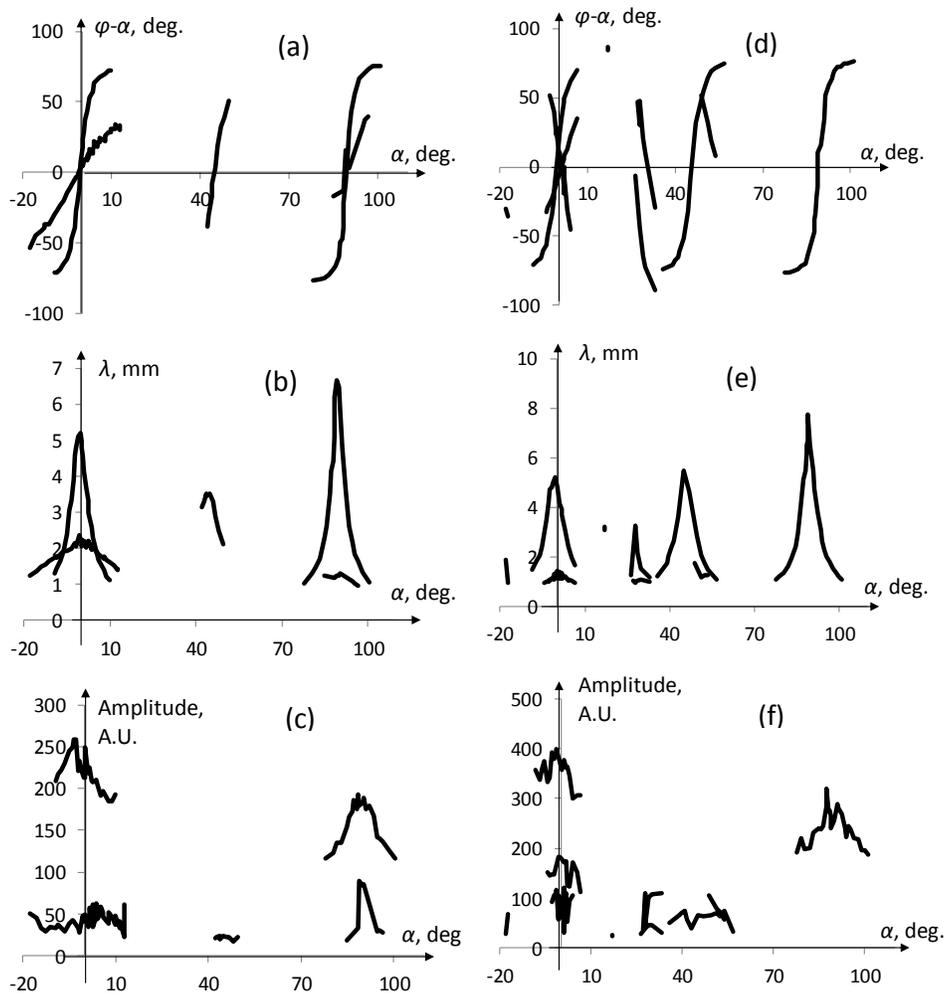

Fig. 11. Three measured parameters of the moiré patterns: orientation (wrapped angle) in the upper row, period in the middle row, and amplitude in the bottom row; plate 50 lpi combined with the plate of different openings: 30% in the left column (a)-(c) and 10% in the right column (d)-(f). Sometimes, especially near the angle 0, two or three branches are detected.

Figure 11 shows that the three functions with longest periods nearly coincide at the angles of 0 and 90°.

It can be noted in Fig. 11 that the experimental amplitude function is similar to the period function. See also Figs. 3(a) and 5, which are centered at the moiré angle. This confirms our assumption [20] about the better visibility of the moiré waves with the longer period. Roughly speaking, the longer is the period of a branch, the higher is the amplitude.

In the upper row of Fig. 11, the moiré waves were detected near the angles of 0, 45°, and 90° (the rational angles 0, 1, and ∞). There is only one wave at the angle of 45°; but two branches were detected at the angles of 0 and 90°. For details, the pattern at the angle 0 has two branches in Fig. 11(a)-(c) and three in Fig. 11(d)-(f); the angle 90° has two branches in Fig. 11(a)-(c) and a single branch in Fig. 11(d)-(f). However, there are a few extra branches with lower amplitudes at the angles other than abovementioned 0, 45°, and 90°. Particularly, there are four additional waves at ±18.4°, 26.7°, 33°, and 56.3° in the lower row of Fig. 11. These low-amplitude branches include fewer points, thus may look somehow incomplete, but still recognizable, as in Fig. 11(d).

For a broader view of the problem, eight moiré angles 0, ±18.4°, 26.7°, 33°, 45°, 56.3°, 63.4°, and 90° (tangents 0, 1/3, ½, 2/3, 1, 3/2, 2, and ∞) were detected by the image processing software in all experiments. We do not distinguish between the positive and negative angles, so we consider the angles -18°, +18°, and 108° as the same rational angle of 1/3. However, the expected angle of 71.6° (tangent 3) was not surely detected, probably because of the small amplitude.

Among all detected moiré angles, the highest amplitudes of the moiré patterns were definitely observed at the angles of 0, 45°, and 90°, see Fig. 12. The average value of the maximum amplitude at 0, 45°, and 90° is approx. 5 times higher than at any other moiré angle.

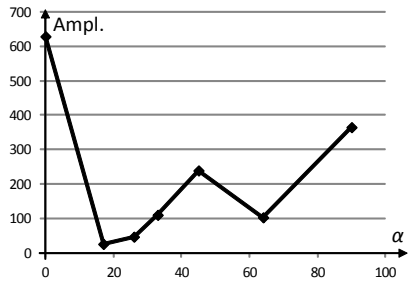

Fig. 12. Maximum amplitude vs. angle.

Across different barrier periods, the period of the moiré patterns is longest, when the period is 0.508 mm (50 lpi), where it is approx. 2.5 times higher than at any other period; while the amplitude of the moiré patterns is highest at 0.339 mm (75 lpi), where it is approx. 3 times longer. The example graphs are shown in Fig. 13. The same or very similar scenario takes place at the angles of 0 and 90°.

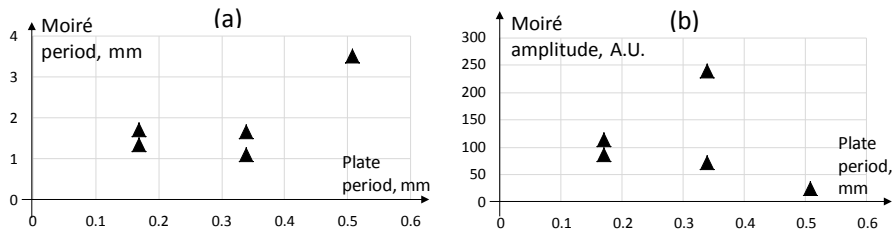

Fig. 13. Period and amplitude of the moiré patterns vs. period of plate at 45°.

The experiments show that the relatively medium periods of the barrier plates (0.3 - 0.5 mm) exhibit the highest amplitudes and the longest periods of the moiré patterns. The angles of 0, 45°, and 90° definitely prevail by their moiré amplitude and period.

Also, we present additional experimental results on the amplitude at 3 angles (0, 27°, and 45°) in 2 displays, see Fig. 14. (The difference of the periods of pixels in both devices is relatively small, about 10% only).

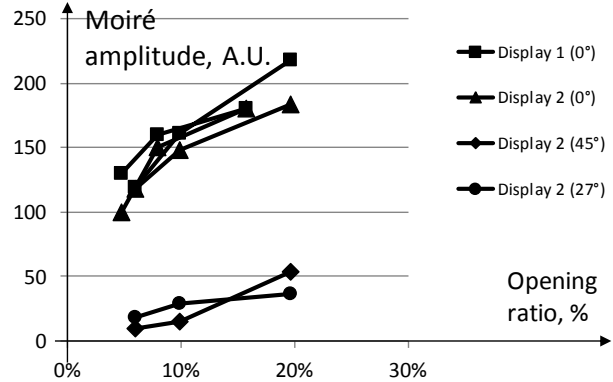

Fig. 14. Amplitude and period of the moiré patterns vs, opening ratio at different angles in two display devices.

Figure 14 confirms that the amplitude of the moiré patterns increases with the opening ratio increased. The average brightness is also increased. Previously, this was shown experimentally [13] at the angle 0 only.

### 4. Discussion

It can be observed in Fig. 11 that the experimental orientation function is clearly S-shaped as predicted by Eq. (4), as well as the experimental period function is bell-shaped by Eq. (5); see also Fig. 3. For the triangular shape of the amplitude function, refer to Eq. (12) and Fig. 5. Among three functions, the measured orientation and the period are clean and smooth, while the shape of the amplitude function looks unclear sometimes. Therefore, the accuracy of the amplitude measurements should be improved.

When the ratio of periods $\rho$ is equal to 1, the maximum period of the moiré patterns is theoretically infinite. However, there is an upper limit for the observed period; it cannot be larger than the size of the screen, where we observe the patterns. Practically, extremely long waves could only be observed as flat, uniform distributions across the whole screen, but cannot be qualified as periodical patterns at all.

The condition, when the moiré patterns are parallel to the lines of the barrier plate, can be a practical indicator of the exact moiré maximum angle.

Our measurement software with confidence recognizes the rational moiré angles up to $m$, $n = 3$. There was only one exception of the missed rational angle (the rational angle 3); which was probably caused by very low amplitude of the moiré patterns at that angle.

Theoretically, the orientation, period, and amplitude functions must be identical in the orthogonal directions 0 and 90°. Assuming the square pixels, we expected the experimental functions to coincide at the angles of 0 and 90°. The experimental Fig. 11 shows that the similarities between the angles 0 and 90° include the size and the shape of functions (there is a difference in the height of the functions about 25% - 30%, though). However the moiré effect at the intermediate angles varies essentially (various sizes and shapes of all functions, sometimes even the absence or presence of the patterns). Also, the failure to detect the expected moiré angle (mentioned tangent 3) confirms that the layout of pixels is not completely symmetric for 0 and 90°.

At the same time, we have to mention that the period of the RGB sub-pixels is equal to the period of the whole square pixels, despite the width of sub-pixels is three times narrower than that of the whole pixel. Correspondingly, the ratio of periods is the same.

The patterns of the higher amplitude and the longer period are better visible. Therefore, the medium periods are most critical in respect to the moiré effect. It means that basically, the most essential impact on the overall picture of the moiré effect is from the period of the grating, while the opening ratio only affects some few particular features like weaker visible branches of smaller amplitudes.

Interestingly, the experimental graph Fig. 12 is somewhat similar to the moiré probability graph in Ref. [11]. This may mean that the probability depends on the amplitude. Therefore, we mostly describe experiments at three main angles: 0, 45°, and 90°, where the moiré effect is strongest. In practice, other rational angles should not be excluded a priori. The angles between the rational angles with $m$, $n \leq 3$ are good candidates for the moiré-free angles in the case of the square pixels in the combination with the linear (1D) barrier.

## 5. Conclusion

The amplitude of the moiré patterns in 3D displays was measured for the first time. The period and orientation of the moiré patterns were also measured as functions of the angle with a small angular increment across a wide angular region. Simultaneous branches are observed and analyzed; the strongest ones selected.

The theoretical interpretation is also given based on the wavevectors, the spectral trajectories, and the MTF. The experimental functions of the period, amplitude, and orientation follow the theoretical predictions.

The proposed method of measurement of the moiré parameters including the amplitude can be useful in comparison of the characteristics of displays based on photographs of the screen with the moiré patterns. It could be one of standard characteristics of the image quality. For example, if one of display devices has a lower moiré amplitude measured under the standard conditions in both displays, its image quality is higher.

The results can be effectively applied to a qualitative comparison of the visual quality of displays, and to the practical selection of a moiré-free angles of the barrier plate and furthermore, to the multi-component (and thus more reliable) minimization of the moiré effect in the digital displays, as well as in other devices where the moiré effect should be reduced or eliminated.